\documentstyle[12pt,aaspp4]{article}
\def\today{\number\day\space\ifcase\month\or
  January\or February\or March\or April\or May\or June\or
  July\or August\or September\or October\or November\or December\fi
  \space\number\year}
\tighten
\journalid{}{}
\articleid{}{}

\slugcomment{Accepted for Publication in the Astronomical Journal}

\begin{document}

\title{THE SPECTACULAR IONIZED INTERSTELLAR MEDIUM \break OF NGC~55} 

\author{Annette M. N. Ferguson\altaffilmark{1} and
 Rosemary F. G. Wyse\altaffilmark{1}}
\affil{Department of Physics and Astronomy, The Johns Hopkins 
University,\break
    Baltimore, MD 21218}
\author{J. S. Gallagher}
\affil{Department of Astronomy, University of Wisconsin, 
Madison, WI 53706}
\altaffiltext{1}{Visiting Astronomer, Cerro Tololo InterAmerican Observatory. 
CTIO is operated by AURA, Inc.\ under contract to the National Science
Foundation.} 
\begin{abstract}

We present deep H$\alpha$+[NII], [SII] ($\lambda\lambda$ 6716,6731 ~\AA)
and [OII] ($\lambda\lambda$ 3726,3729 ~\AA) images of the highly
inclined, actively star--forming SBm galaxy NGC~55, located in the
nearby Sculptor Group.   Due to its proximity, NGC~55 provides a unique
opportunity to study the disk--halo interface in a late--type galaxy
with unprecedented spatial resolution.  Our images reveal a spectacular
variety of ionized gas features, ranging from giant HII region
complexes, to supergiant filamentary and shell features, to patches of
very faint diffuse emission.     Many of these features protrude well
above the plane of the galaxy, including a very faint fragmented shell
of emission which is visible at 2.6~kpc above the disk.  We identify
candidate `chimneys' extending out of the disk, which could be the
conduits into the halo for hot gas around disk star-forming regions,
and could also provide low-density paths for the passage of UV photons
from the disk to the halo.  Several of the identified chimneys are
`capped' with clumps of ionized gas, one of which, located at 1.5~kpc
above the disk plane, appears to be the site of recent star formation.
Emission--line ratios ([OII]/H$\alpha$+[NII], [SII]/H$\alpha$+[NII])
constrain the ionization mechanism of the gas, and our images allow the
first measurement of [OII]/H$\alpha$+[NII] in extra-planar diffuse
ionized gas.  The diffuse gas is characterized by emission--line ratios
which are enhanced on average by a factor of two compared to those of
bright HII regions.  Each line ratio increases in value smoothly from
the cores of HII regions, through the haloes of HII regions, into the
diffuse ionized gas.  Such a continuous trend is predicted by models in
which the diffuse gas is ionized by photons produced by massive stars
in HII regions.  We discuss the factors which may determine the
existence of extra--planar ionized gas in galaxies.

\end{abstract}
\keywords{Galaxies:  Individual (NGC~55) -- Galaxies: Irregular --
Galaxies: ISM -- ISM: bubbles -- ISM: HII regions -- ISM: structure}
\newpage
\section{Introduction}

The interstellar disk--halo connection plays an important role in
understanding the processes of disk galaxy evolution (e.g. Bloemen
1991).  Many theories have been proposed by which star formation in the
disk causes continuous and/or sporadic transfer of mass and energy from
the disk to the halo, with the possible return of halo material to the
disk (e.g. Shapiro \& Field 1976, Norman \& Ikeuchi 1989; hereafter
NI89).  Such a cycle could regulate the star formation process in
disks, leading to low overall star-formation efficiency, as well as
provide a means by which metal--enriched gas could be distributed over
large regions of galaxies.  This disk--halo interaction may also help
in creating and sustaining the ionized gaseous haloes inferred to exist
around galaxies from quasar absorption-line studies (e.g. Bergeron \&
Boisse 1991).

Several pieces of evidence suggest a strong coupling between the disk
and  halo in galaxies.  In the Milky Way,  structures resembling
$`$worms' in the atomic gas have been found,  typically extending
several hundred parsecs out of the disk into the halo (e.g. Heiles
1984; Normandeau, Taylor \& Dewdney 1996) and there exists a thick
ionized gas  layer, the $`$Reynolds layer' with a scale-height of
$\sim$ 1 kpc (Reynolds et al 1971; Reynolds 1984).  Maintaining the
ionization of this layer requires a large power that can be met easily
only by ionizing photons from OB stars  (e.g. Reynolds 1984; Kulkarni
and Heiles 1988) however these photons must be able to travel from
their point of origin in the disk to significant heights above the
plane.  Other possibilities for the origin of the ionizing photons have
been proposed and include shocks, turbulent mixing layers (Slavin,
Shull \& Begelman 1993) and galactic microflares (Raymond 1992).  In
recent years, deep narrow--band imaging of  highly inclined external
disk galaxies has revealed that some form of  diffuse ionized gas, a
term commonly used to refer to ionized gas lying outside the boundaries
of traditional HII regions, is quite often found above the disk plane
(e.g. Rand 1996).  Most often, the extra--planar ionized gas is seen in
the form of discrete structures, such as loops, filaments and plumes,
occasionally accompanied by an apparently unstructured, thick diffuse
layer (e.g.  Rand, Kulkarni \& Hester 1990, 1992; Pildis, Bregman \&
Schombert 1994; Rand 1996).  Diffuse ionized gas has also been observed
in several face--on galaxies (e.g.  Monnet 1971; Hunter \& Gallagher
1990, 1997; Walterbos \& Braun 1994; Ferguson et al 1996),  where a
clear correlation with sites of very recent star formation has
emerged.

Local energy and momentum input, and local mass surface density, govern
the disk--halo connection (e.g. NI89, Heiles 1990).  Only the most
luminous OB associations are likely to produce superbubbles which can
\lq break--out' of the disk, and thus relatively late--type systems,
which typically host the most luminous HII regions (Kennicutt 1988),
are of extreme interest.  Low mass, low luminosity galaxies, such as
the LMC are observed to host several supershells and giant filaments
(e.g. Davies, Elliot \& Meaburn 1976, Hunter, Hawley \& Gallagher 1993)
but the vertical extent of these structures is unknown.

We here present a study of the disk-halo interface in the nearby highly
inclined SBm galaxy NGC~55,  located in the nearby Sculptor Group.  Our
study is based on large field--of--view H$\alpha$+[NII], [SII]
($\lambda\lambda$ 6717,6731 ~\AA) and [OII] ($\lambda\lambda$ 3727,3729
~\AA)  images.  Due to its proximity, NGC~55 provides a unique
opportunity to study the disk--halo interface in a late--type galaxy
with unprecedented spatial resolution.  The basic properties of the
galaxy are listed in Table 1.  Hoopes et al (1996) have recently
analysed H$\alpha$+[NII] and [SII] images of NGC~55, taken with the
CTIO Schmidt telescope.  The data presented here are of considerably
higher spatial resolution and sensitivity, and include images in the
important [OII] emission line, allowing a more detailed study of
structure and ionization state of the gas in the disk--halo interface.

\section{Observations and Data Reduction}

\setcounter{footnote}{0}

Images of NGC~55 were obtained using the CTIO 1.5~m + Tek 2048 x 2048
CCD during September 1994 and 1995.  At f/7.5, the pixel scale was
0.44$''$, corresponding to physical size of $\sim 4$~pc at the distance
adopted here for NGC~55 of 1.6~Mpc.   The filters used for the
observations were well--matched to the galaxy's recessional velocity
and consisted of the following: (i) 68~\AA (FWHM) H$\alpha$ filter
(which also encompassed the [NII](6548,6584~\AA) lines) and a broadband
R filter for the continuum observations, (ii) 36~\AA (FWHM) [SII] filter
and a narrow (80~\AA) off-band filter for the continuum observations,
and (iii) 24~\AA (FWHM) [OII] filter and a narrow (130~\AA) off--band
filter for the continuum observations. The field of view was $\sim$ 15
$\times$ 15 arcmin$^2$ (corresponding to 7~kpc on a side at the
distance of NGC~55)  for the H$\alpha$+[NII] observations and slightly
smaller than this for the [OII] and [SII] filters, due to vignetting.
Complete spatial coverage of the galaxy was achieved in H$\alpha$ by
three separate pointings; the [SII] and [OII] observations were
restricted to the central 5~kpc.  Total exposure times were
approximately 2 hours in H$\alpha$ (per field) and 1.5 hours each in
[SII] and [OII].  The seeing was in the range of 1.5 -- 2.0 $''$.

Reductions were carried out using standard procedures.  Well--exposed,
median--filtered twilight sky frames were used to flat--field the
images, and an accuracy of better than 1\% was typically achieved.  Multiple
pointings at the approximately the same  position were registered using
field stars in the frame and those taken through the same filter were
combined with an average sigma--clipping technique.   The sky value was
determined in each image as the mean of the median pixel value in a
series of small 100$\times$100 pixel boxes placed well outside the
galaxy, and subtracted off.   Foreground stars in the frames were used
to determine the scaling factors between the emission--line and the
continuum images.   The scaled continuum images were subsequently
subtracted to produce net emission--line images.

Observations of standard stars from the list of Stone and Baldwin
(1983) were used to calibrate the H$\alpha$+[NII] and [SII] images.
Our NGC~55 H$\alpha$ frames  were largely obtained under
non--photometric conditions; the few frames taken under photometric
conditions were calibrated using standard stars, then those frames were
used to calibrate our final combined images.  The [SII] images were
obtained under photometric conditions.   The [OII] image was calibrated
relative to the H$\alpha$ frame using published spectrophotometric
data for several HII regions in the central region of the galaxy
(Webster \& Smith 1983).   The average sensitivity of the
H$\alpha$+[NII] continuum--subtracted image, taken to be 1$\sigma$ of
the sky background, is 1.59$\times 10^{-17}$~erg~
s$^{-1}$~cm$^{-2}$~arcsec$^{-2}$, corresponding to an emission measure
{\footnote{Emission measure is related to Rayleighs, the commonly used
unit of surface brightness in Galactic DIG studies, by EM (pc
cm$^{-6}$) = 2.78$\times$I$_{H\alpha}$ (Rayleighs) for T$_{e}$=10$^4$
K.} {\it per pixel} of 7.8~pc~cm$^{-6}$ (for an assumed electron
temperature of 10$^4$K). When smoothed with a gaussian of FWHM =
2.5$''$, to enhance the detection of very faint, extended features, the
sky noise decreases to EM~$\sim$~1~pc~cm$^{-6}$.  The average
sensitivities of the unsmoothed [OII] and [SII] images are 4.82$\times
10^{-17}$~erg~ s$^{-1}$~cm$^{-2}$~arcsec$^{-2}$  and 1.32$\times
10^{-17}$~erg~ s$^{-1}$~cm$^{-2}$~arcsec$^{-2}$ respectively.  The
contribution of [NII] to our H$\alpha$+[NII] images is expected to be
small,  based on the published spectra of HII regions from Webster \&
Smith (1983).  Galactic extinction is negligible in the direction of
NGC~55 and no corrections have been applied.

Scattered light is an important issue in studies of faint extended
emission, and  can be caused either by the telescope optics/detector
or, in the particular case of ionized gas, by dust in and around HII
regions.  Scattering in the optics/detector of the system can be
constrained by measuring the azimuthally averaged radial light profile
around several bright stars in our fields. From this analysis, we find
that 95\% of the total light from a star is contained within a radius
of $\sim$~6{\arcsec} in the H$\alpha$+[NII] image, and within a radius
of $\sim$~5{\arcsec} in the [SII] and [OII] images.  Furthermore, in
all three images, more than 99\% of the total light from a star is
enclosed within a radius of $\sim$ 20{\arcsec}.   Thus, if the faint
diffuse ionized gas emission present in NGC~55 were simply a result of
scattered light in the telescope, then it should be significantly more
localised around HII regions than is observed, and should contribute no
more than a few percent of the total emission--line flux measured in
each filter.  Scattering from dust in the HII regions themselves is
unlikely given the distinct morphology of the diffuse emission, as well
as the different emission--line ratios of the diffuse emission relative
to HII regions. 
 
\section{ Morphology of the Ionized Gas}

Figure~1 (Plate X) shows our H$\alpha$+[NII] continuum--subtracted
mosaiced image of the entire galaxy.   A spectacular variety of ionized
gas features are visible, ranging from bright HII complexes, to loops
and plumes protruding out the plane of the galaxy, to faint diffuse
emission lying both above, below and within the disk.  These features
are detected in all three bands, however filamentary features are less
pronounced in the [OII] and [SII] images due their lower S/N.  The
complex structure seen in the extra--planar ionized gas is striking.
Our images underscore the importance of high spatial resolution in
studying the structure of the disk--halo interface in galaxies; pending
HST observations of NGC~891 (Dettmar, private communication), in which
the extra-planar gas studied thus far does {\it not} show such detailed
structure, will probe a similar spatial scale as the NGC~55
observations presented here.

The most prominent features are the two large HII complexes located in
the central regions of NGC~55, which are clearly the dominant sites of
recent massive star formation.  The H$\alpha$+[NII] luminosities of
these features are estimated to be 9  and 3 $\times$ 10$^{39}$
erg~s$^{-1}$ (uncorrected for extinction) and they each span more than
a kpc in diameter.   Their individual luminosities are comparable to
that of the 30 Dor complex in the LMC (Kennicutt et al 1995) and are
more slightly luminous than NGC~604, the brightest HII complex in M33
(Kennicutt 1988).    The combined luminosity of these two complexes
accounts for more than half of the total H$\alpha$+[NII] luminosity of
NGC~55, and exceeds that observed for the entire ensemble of HII
regions in the disk of M31 (Walterbos \& Braun 1994).

These two HII complexes are surrounded by a wealth of 
structure.  Figure 2~a (Plate~XX) shows an H$\alpha$+[NII] image of the
central $\sim$ 5~kpc of the galaxy, with many of these features marked;
Table~2 lists their measured properties.  Several of the
brighter features have been previously discussed:  Graham \& Lawrie
(1982) identified feature $`$D', during a photographic plate search for
planetary nebulae; Bomans \& Grebel (1993) identified features  $`$F'
and $`$H' from a small field of view CCD H$\alpha$ image centered on
the two brightest HII regions. Hoopes et al.  (1996) have discussed the
general morphology of the ionized gas emission from shallower, lower
resolution images obtained with the CTIO Schmidt Telescope.

Several of the ionized gas features in NGC~55 extend considerable
distances above the disk plane. Features $`$B', $`$E', $`$G', $`$I' and
$`$K' may be examples of chimney structures, through which hot gas is
theorised to be vented into the halo.  Such structures are predicted to
arise as a result of correlated Type II SNe explosions in OB
associations, leading to the production of an expanding superbubble
which can break through the disk (e.g. MacLow, McCray \& Norman 1988;
Norman \& Ikeuchi 1989; Heiles 1990).  In our images, several of these
features ($`$G', $`$I' and $`$K') appear to be capped with clumps of
relatively bright emission, possibly from gas swept up by the expanding
bubble and now directly ionized by the OB association located in the
disk below (see Figure 2b).  The cap which sits atop feature $`$I' is
particularly interesting, since it is strong in the [OII]--continuum
image and may actually be due to {\it in situ\/} star formation at
1.5~kpc above the plane of the disk.   It has an H$\alpha$+[NII]
luminosity of $\sim$ 3~$\times$~10$^{36}$~erg~s$^{-1}$, slightly less
luminous than that of the Orion nebula (Kennicutt 1984).  On the North
side of the galaxy, we have detected several patches of extremely faint
H$\alpha$ emission (EM $<$ 5 pc cm$^{-6}$) lying at distances up to
2.6~kpc above the plane.  These patches, most clearly seen on Figure
2~b (Plate XXX), which has been smoothed to enhance the detection of
features at faint levels, are also visible in our [SII] and [OII]
images.  These clumps of faint emission appear to trace out a
fragmented shell, which has possibly undergone Raleigh--Taylor
instability (see review of Tenorio--Tagle \& Bodenheimer 1990).
 
Diffuse ionized gas emission can be seen throughout the disk of the
galaxy, however it appears brighter and more structured near
concentrations of young, massive stars.   Even the more quiescently
star--forming outer parts of the galaxy appear to have significant
quantities of faint diffuse ionized gas, although in these parts it is
largely confined within the plane of the galaxy.  The small--scale
correlation between diffuse ionized gas and HII regions has been
observed in several galaxies (e.g. Walterbos \& Braun 1994, Ferguson et
al. 1996) and indeed in NGC~55 (Hoopes et al 1996)  and lends support
for a model in which the gas is photoionized, either directly or
indirectly, by Lyman continuum photons which leak out of individual HII
regions.  Qualitatively, it also appears that the {\it intensity} of local
star formation (ie. SFR per unit area) plays a role in determining the
surface brightness and the morphology of the surrounding diffuse ionized gas.

There is some evidence for a thick diffuse ionized gas layer in the
central regions of NGC~55.  Averaged over the central 1.4~kpc, the
full--width of this layer (in H$\alpha$+[NII]) at an EM $=$ 10 pc
cm$^{-6}$ is  $\sim$ 2~kpc.  This extended emission is also seen in the
[OII] and [SII] images, where the layer has a full--width of
$\sim$~2.6~kpc and $\sim$~1.4~kpc respectively, at the aforementioned
surface brightness level.  Uncertainties in flat--fielding and the
possibility of residual instrumental effects make the detection of this
layer marginal at present.  Future observations  will allow us to study
this extra--planar ionized gas layer in more detail.

\section{Luminosity of the Ionized Gas}

The global H$\alpha$+[NII] luminosity of NGC~55 is found by summing all
pixel values within a large polygonal aperture.  We derive a value of
2.0 $\times$ 10$^{40}$~erg~s$^{-1}$; this value is about 20\% lower
than that recently reported by Hoopes et al (1996).   Large scale
flat--fielding errors, as well as realistic errors in the continuum
scaling factor (estimated to be $\lesssim$ 3\%) could produce a
systematic uncertainty of $\pm$ 5--10\% in this measurement.  Assuming
an IMF and a set of stellar evolutionary tracks, the global
H$\alpha$ luminosity can be translated into a global current star
formation rate.   Kennicutt, Tamblyn \& Congdon (1995) have calculated
the proportionality  factor between the two for various IMFs, based on
the stellar evolutionary tracks of Schaller et al (1993).   Using these
calculations for convenience, the global star formation rate of NGC~55,
over the mass range 0.1~M$_{\sun} \leq$~M$ \leq$ 100~M$_{\sun}$, is
found to be 0.16~M$_{\sun}$ yr$^{-1}$, assuming a Salpeter IMF, or
0.15~M$_{\sun}$ yr$^{-1}$, assuming a modified Miller--Scalo IMF
(Kennicutt 1983).  These numbers are likely to be upper limits to the
true current star formation rate since we expect to have a small contribution
from [NII] in our global H$\alpha$ flux.

Isolating the fraction of the total H$\alpha$ emission which is
produced by diffuse ionized gas is a difficult task, and many different
approaches have previously been adopted (e.g. Walterbos \& Braun 1994,
Veilleux et al.  1995, Ferguson et al. 1996, Hoopes et al. 1996).   In
view of the many uncertainties inherent in deriving the diffuse
fraction, especially in the case of a highly inclined galaxy (e.g.
variable dust extinction in the plane), we choose here to adopt the
simplest technique. An isophotal cut in surface brightness was imposed
at EM$=80$~pc~cm$^{-6}$, corresponding to the limit below which the
H$\alpha$ emission could largely be classified as filamentary and/or
diffuse. We derive a global diffuse fraction in H$\alpha$+[NII] of 19\%
$\pm$ 5\%,  which can be compared with the value of $\gtrsim 30$\%
estimated by Hoopes et al. (1996) using a different technique (the
uncertainty associated with our measurement is estimated from realistic
errors in the large--scale flat--fielding and continuum subtraction).
These same authors also present a diffuse fraction measured via an
isophotal cut in surface brightness; at our adopted value of 80 pc
cm$^{-6}$, they find a diffuse fraction of 26\%, being slightly higher
than the value found here.

Luminosities in H$\alpha$+[NII], [OII] and [SII] were calculated for
the central 5~kpc of the galaxy and are 1.3$\times$
10$^{40}$~erg~s$^{-1}$, 1.2$\times$ 10$^{40}$~erg~s$^{-1}$ and
3.4$\times$ 10$^{39}$~erg~s$^{-1}$ respectively, uncorrected for
extinction.  Systematic uncertainties on these values due to flat--fielding and
continuum subtraction errors are estimated to be $\lesssim$ 10\%.  We
used the H$\alpha$+[NII] image to define a mask in which pixels with EM
$\leq$~80 ~pc~cm$^{-6}$ were replaced with a value of 0, and those
above this limit were replaced with 1.  Multiplying each of the
emission--line images by this mask produced images containing
predominantly HII region emission.  Summing the counts in these images,
and subtracting them from the total counts measured on the unmasked
images, allowed us to compute the diffuse fractions in the different
bands, for the central 5~kpc of the galaxy.  Diffuse fractions of
17\% $\pm$ 3\% were derived in H$\alpha$+[NII], 29\% $\pm$ 5\%
in [OII] and 26\% $\pm$ 5\% in [SII].

The total and diffuse ionized gas luminosities in each of the emission
lines enabled us to calculate the mean line ratios
([OII]/H$\alpha$+[NII] and [SII]/H$\alpha$+[NII]) for the discrete HII
regions and the diffuse ionized gas.  We find [OII]/H{$\alpha$}+[NII]
$\sim$ 0.80 and [SII]/H$\alpha$+[NII] $\sim$ 0.24 for the HII regions,
and [OII]/H$\alpha$+[NII] $\sim$ 1.42 and [SII]/H$\alpha$+[NII]
$\sim$~0.43 for the diffuse ionized gas.  We will investigate this
apparent difference in ionization state between the HII regions and the
diffuse ionized gas further in the following section.

\section{Line Ratios and Ionization Mechanism of the Gas}

Constraints on the ionization mechanism of the extra--planar gas are
provided from our data by the values of the line ratios of
[OII]/H$\alpha$+[NII] and [SII]/H$\alpha$+[NII].  Theories which
explain the diffuse ionized gas as being a result of photoionization by
a dilute radiation field (e.g.  Mathis 1986, Domg\"orgen \& Mathis
1994)  predict these forbidden line ratios should be significantly
enhanced in the diffuse ionized gas relative to bright HII regions, and
that there should be a smooth trend of increasing ratio with increasing
distance from the ionizing source.  Predictions from shock-ionization models,
particularly for the [OII] line strengths,  are rather strongly
dependent on parameters such as the shock velocity (Shull \& McKee
1979; Raymond 1980), which may be highly variable, from one location to
another.

Our data allow us to study the variation of emission line ratios over wide
ranges of surface brightnesses and of morphology, from bright HII
region cores to faint DIG features.  Small aperture (2$''$) photometry
was carried out at different locations along the features identified in
Figures 2~a as well as in the core and halo regions of several bright
HII regions.  The total flux inside a given aperture was measured
separately on each of the H$\alpha$+[NII], [OII] and [SII] images and
then ratioed.  Only those measurements in which the S/N $\gtrsim$ 3 in
both apertures were retained; this eliminated $\sim$ 10\% of the
measurements.  

It is important to assess the possible effects of internal extinction 
on the derived emission line ratios.
Internal extinction within NGC~55 should not affect the
[SII]/H$\alpha$+[NII] ratios measured in various features, where the
lines are close in wavelength, but may be an important factor in
comparing the [OII]/H$\alpha$+[NII] ratios.  The possible amplitude of
this may be estimated using the published spectrophotometry  of Webster
\& Smith (1983), which yields a mean extinction at H$\beta$ of
C(H$\beta$)=0.35 for the cores of bright HII regions (ie. EM $\sim$ 10$^5$
pc cm$^{-6}$) in the disk of the NGC~55.  Adopting the  LMC reddening
law  of Howarth (1983), we find that the intrinsic [OII]/H$\alpha$
ratio towards these positions could be a factor  of $\sim 1.7$ higher
than that observed (this value is only slightly lower if the Milky Way
extinction law is adopted instead).  If the haloes surrounding the HII
regions and the diffuse ionized features located close to the plane of
the galaxy were extincted to the same extent as the HII region cores,
then one would expect a similar effect.  We have no direct knowledge
the distribution of dust across, or perpendicular to, the disk of
NGC~55 however; the two central HII regions studied by Webster \& Smith
(1983) show variations in C(H$\beta$) of order 0.2, indicating that the
dust content in these regions is clearly patchy.  In view of these
uncertainties,  we have not attempted to correct our line ratios for
internal extinction.  The possible influence of dust on the observed
measurements should be kept in mind in the following discussion,
however it will not change our major conclusions.

\subsection{Trends with H$\alpha$ Surface Brightness}

Figures 3~a \& b show the measured emission line ratios plotted against
H$\alpha$ surface brightness.  There is a clear increase in both the
[OII]/H$\alpha$+[NII]  and [SII]/H$\alpha$+[NII] ratios as the
H$\alpha$+[NII] surface brightness of the measured feature (within the
aperture) decreases. [OII]/H$\alpha$+[NII] ratios of 0.2 -- 1.2 are
typically found for the HII regions (a mean value of 0.4 is found for
the bright cores), whereas the diffuse ionized features have ratios in
the range 0.7 -- 2.5; likewise, [SII]/H$\alpha$+[NII] ratios of 0.05 --
0.3 typify the HII regions (a mean value of 0.08 is observed in the
cores) whereas values of 0.2 -- 0.7  are found in the diffuse gas.
Systematic uncertainties in the emission--line ratios due to realistic
errors in the continuum subtraction are typically $\pm$ 0.01 for bright
HII regions and $\pm$ 0.05 for diffuse ionized features.  The mean
[OII]/H$\alpha$+[NII]  and [SII]/H$\alpha$+[NII] ratios measured for
the diffuse ionized features are found to be a factor of $\sim$ 2
higher than the mean found towards HII regions, and factors of 4--5
higher than those found in the bright cores of HII regions; this is in
agreement with our earlier calculation based on mean luminosities.
[SII]/H$\alpha$ ratios in this general range have been observed for
diffuse ionized gas in the Milky Way and other galaxies and several
authors have pointed out the enhancement relative to bright HII regions
(e.g. Sivan, Stasinska and Lequeux 1986; Reynolds 1985, Walterbos \&
Braun 1994, Hunter 1994; Hoopes et al 1996).  Furthermore, Hunter
(1984, 1994, 1996) has found enhanced [OII]/[OIII] compared to bright HII
regions in many of the ionized filaments and shells in the LMC and
other late--type systems.

\begin{figure}[h]
\figurenum{3}
\plottwo{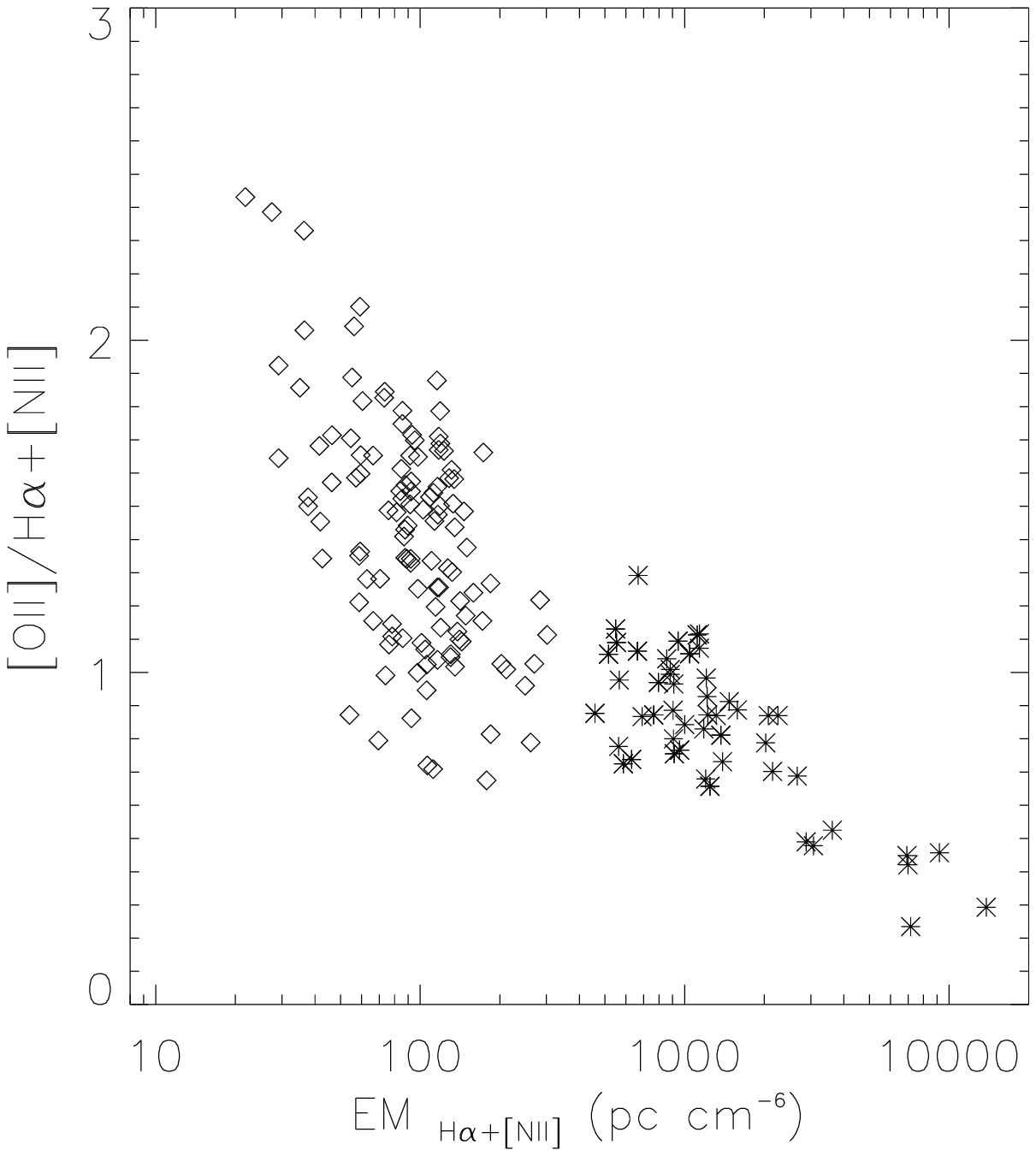}{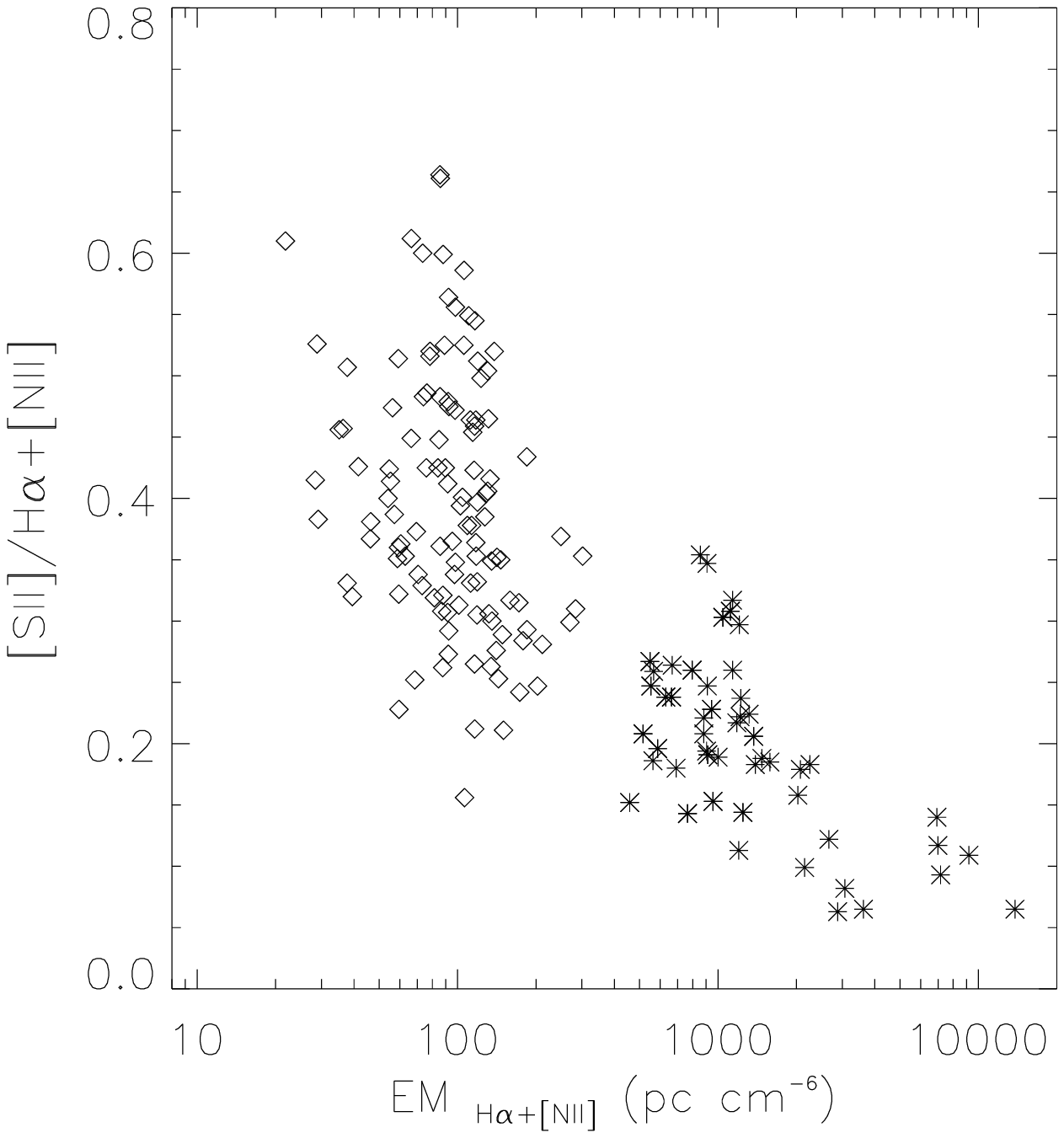}
\caption{ (a)~The variation of [OII]/H$\alpha$+[NII] vs
H$\alpha$+[NII] surface brightness for various HII regions and diffuse
ionized gas features in NGC~55.  HII regions are marked by ($\ast$) and
diffuse ionized gas features by ($\Diamond$).  (b) ~~ The variation of
[SII]/H$\alpha$+[NII] vs H$\alpha$+[NII] surface brightness for various
HII regions and diffuse ionized gas features in NGC~55.}
\end{figure}

Figures 3~a \& b also reveal the presence of a smooth trend in
emission line ratio as one proceeds from bright HII region cores,
through haloes of HII regions and into the diffuse ionized gas.  HII
regions and diffuse ionized gas features observed at the same
H$\alpha$+[NII] surface brightness have line ratios which are very
similar.  Such a continuous trend between the faint outskirts of HII
regions and diffuse ionized gas was not found in the  Milky Way, where
a distinct difference between the two is observed to exist (Reynolds
1985, 1988).  A very marginal trend in [SII]/H$\alpha$+[NII] vs
H$\alpha$ surface brightness was seen by Walterbros \& Braun (1994) but
was not seen by Hoopes et al (1996) in their study of three Sculptor
Group galaxies (including NGC~55).  This could be related to the larger
apertures these authors used for their aperture photometry, and the
lower S/N of their data.  Spectroscopic study of several filaments and
supershells in the LMC and NGC~1800 by Hunter (1994, 1996) found
evidence for a correlation between [OII]/[OIII] and H$\alpha$ surface
brightness, in the sense that lower surface brightness features were
characterized by a lower ionization state.   The results presented here
can be similarily interpreted.

Finally, while some scatter does exist in these plots, different
morphological classes of diffuse ionized gas structures (e.g. loops,
chimneys, patches of diffuse emission) cannot be distinguished from
each other purely on the basis of the line ratios presented here.  This
has also been observed by Hunter (1994, 1996) who studied different
classes of ionized structures in the LMC and NGC~1800.  This is
consistent with all having a common ionization mechanism, which as we
discuss in Section 6, is most likely to be photoionization by massive stars.

\subsection{Trends with Height above the Plane}

Figures 4a \& b show the behavior of the emission line ratios with
height above the plane.   A clear trend is seen in the
[OII]/H$\alpha$+[NII] ratio, amounting to a factor of $\sim$~4
difference between features in the plane and those above it.   The
[SII]/H$\alpha$+[NII] ratio also shows a trend with height above the
plane, although there is more scatter.   This seems to be due  both to
low H$\alpha$+[NII] surface brightness features in the plane having a
high [SII]/H$\alpha$+[NII] ratio,  as well as some bright extra--planar
features which have low [SII]/H$\alpha$+[NII] ratios.   For example,
the cap of emission which sits atop feature $`$I', and the large shell
(feature $`$D') have ionization states more typical of HII regions than
diffuse ionized features ($`$I': [OII]/H$\alpha$+[NII] $\sim$ 1,
[SII]/H$\alpha$+[NII] $\sim$ 0.3, $`$D': [OII]/H$\alpha$+[NII] $\sim$
1.5, [SII]/H$\alpha$+[NII] $\sim$ 0.3).  Graham \& Lawrie (1982)  also
observed their shell to have unusually strong [OIII] emission
([OIII]/H$\beta$ $\sim$ 1).  These line ratios could be understood if
these structures were being directly ionized by  OB associations in the
disk below, or by {\it in situ\/} star formation.  It should be noted
that enhanced [NII]/H$\alpha$ ratios have been observed in the
extra--planar diffuse ionized gas in NGC~891 (Dettmar \& Schulz 1992)
and NGC~4631 (Golla et al 1996); if such an enhancement is also present
in the diffuse ionized gas of NGC~55, then this implies that the true
[SII]/H$\alpha$, [OII]/H$\alpha$ line ratios in the extra--planar gas
could be even higher than presented here.

\begin{figure}[h]
\figurenum{4}
\plottwo{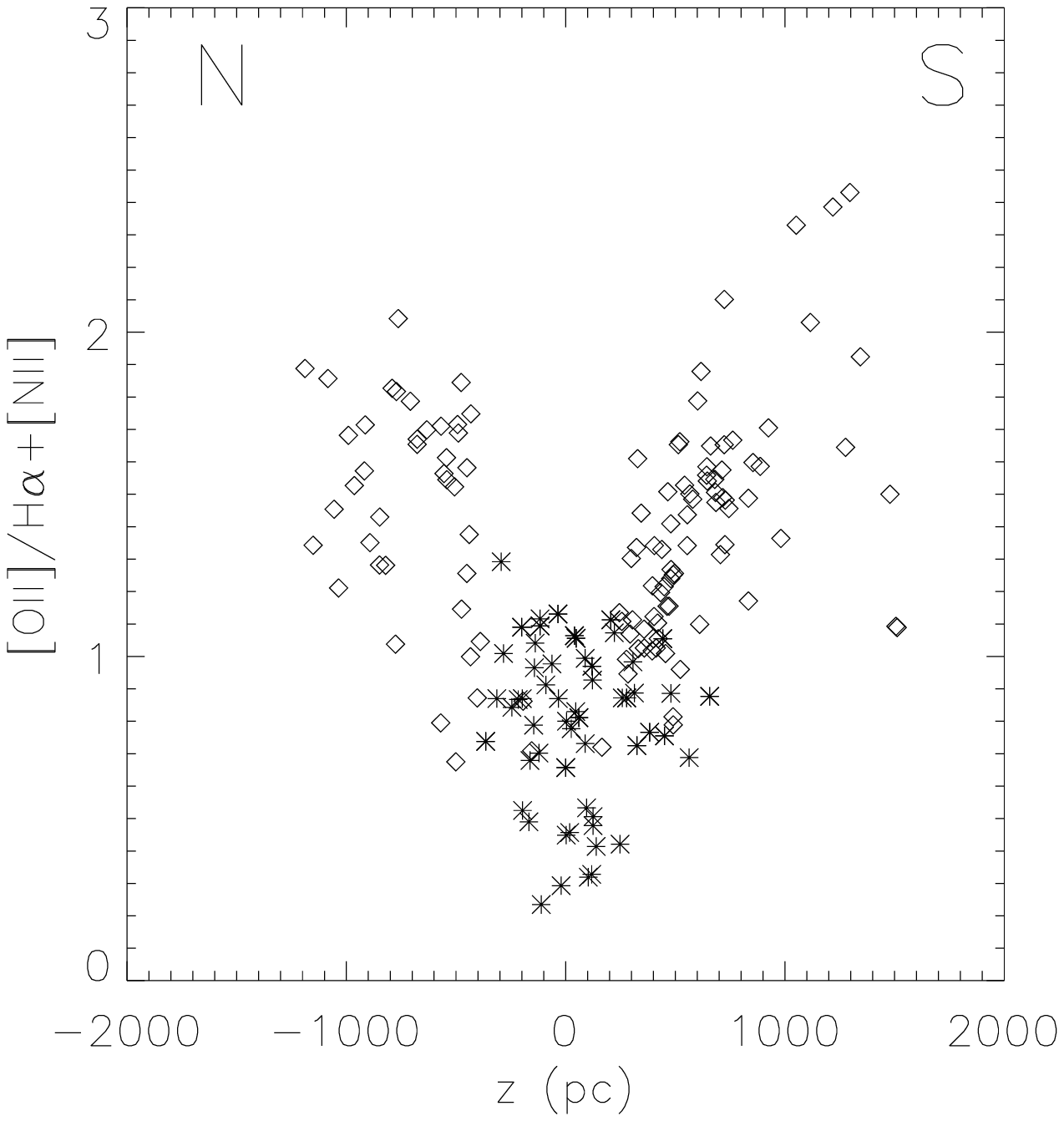}{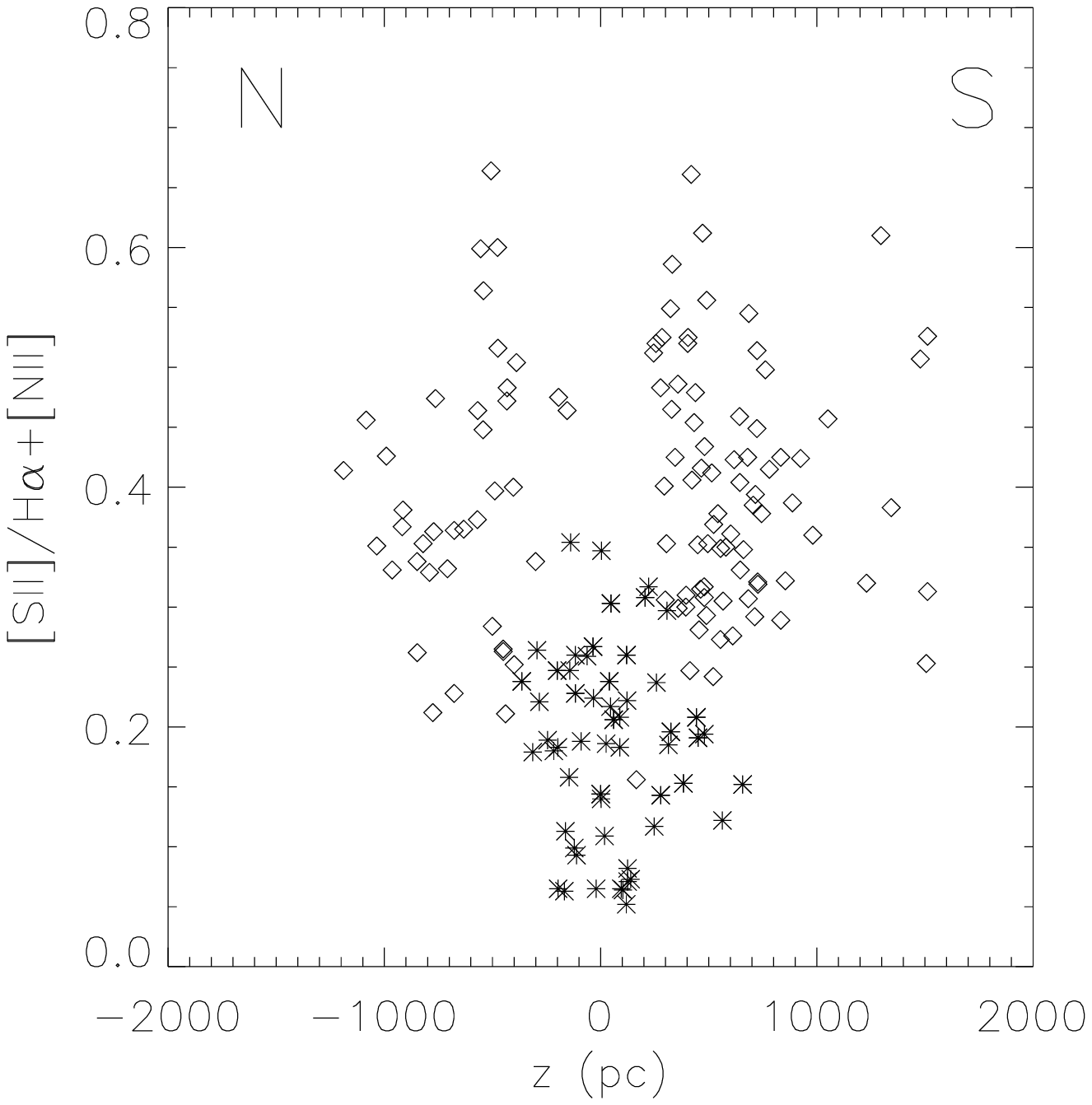}
\caption{(a) ~~The
variation of [OII]/H$\alpha$+[NII] vs z (height above the plane in
parsecs) in NGC~55. HII regions are marked by ($\ast$) and
diffuse ionized gas features by ($\Diamond$). North and South are labelled.  (b) The variation of
[SII]/H$\alpha$+[NII] vs z (height above the plane in parsecs).}
\end{figure}

The lack of a well-defined correlation between [SII]/H$\alpha$+[NII]
and $z$ at first appears to be in contrast to the studies of Rand et al
(1990), Dettmar \& Schulz (1992) and Golla et al (1996) where a clear
trend was seen in the edge--on galaxies NGC~891 and NGC~4631.  However,
those trends were seen in single vertical cuts taken parallel to the
minor axis, intersecting the disk at points of particularly bright HII
region emission.  Our plots on the other hand show the trend for a
variety of disk features and include faint diffuse in disk emission, as
well as bright extra--planar filaments and loops.  Indeed, experiments
with our data have revealed that a single cut through the bright
central HII complex shows a significantly stronger correlation between
[SII]/H$\alpha$+[NII] and $z$ than that seen in Figure 4b.

\section{Discussion}

We have presented a deep H$\alpha$+[NII], [SII] and [OII] imaging study
of the nearby, highly inclined galaxy NGC~55, allowing us to probe the
disk--halo interface with unprecedented spatial resolution.  NGC~55
hosts a spectacular variety of ionized gas features,  protruding
typically 1--2~kpc out of the plane of the disk, as well as what
appears to be a faint diffuse ionized layer extending over the central
regions.  The numerous chimneys protruding out of the disk may
facilitate the transport of hot gas into the halo and provide low
column-density pathways by which UV photons can travel to high
scaleheights.    A very faint shell of emission is seen which reaches a
height of $\sim$2.6~kpc.  
 
Despite its low luminosity and current star formation rate,  NGC~55 has
a disk--halo interface which rivals that seen in galaxies with higher
global star formation rates, such NGC~891 and NGC~4631 (Rand 1990;
Dettmar 1990).  Most of the extra-planar activity seen in NGC~55 is
concentrated around the two large HII complexes in the central regions,
suggesting that the intensity of local star formation governs both the
morphology and the surface brightness of diffuse emission (see also
Ferguson et al 1996).  Indeed many low mass dwarf and irregular
galaxies dominated by a single large HII complex are observed to host a
variety of large ionized filamentary structures, some may even protrude
out of the plane (e.g.  NGC~1800, NGC~1569), despite the fact that
their global star formation rates are extremely modest (Marlowe et al
1994; Hunter et al 1993).   As expected theoretically, {\it local},
rather than {\it global}, star formation properties are important for
the presence of structured extra--planar ionized gas in disk galaxies.
Smoother diffuse ionized gas layers such as seen in the Milky Way and
in M31, which lack super-giant HII regions, could require simply a
source of ionizing photons, such as leaky HII regions (Leisawitz \&
Hauser 1988; De Geus et al 1993) and a relatively porous ISM  whereas
large extra--planar structures could require more violent star
formation, which structures the ISM and results in more pathways  along
which UV photons can travel unscathed.   Further observations of
edge--on and face--on disk galaxies will allow these ideas to be
tested.

One should bear in mind that the means by which gas is physically
transported to large distances above the plane may well be different
from the process which keeps it ionized.  Many factors  play a role in
the formation of superbubbles and chimneys, including the ambient gas
density and magnetic fields (NI89, Tomisaka 1988, Ferriere et al
1989).  The strong curvature of many of the filaments towards the plane
of the NGC~55 suggests they could shaped by a magnetic field, which is
expected to have this geometry if, for example, the Parker instability
or related processes are occuring (e.g. Parker 1979).  Gas could be
 clumped along the magnetic field lines and simply illuminated by
escaping UV photons, or the magnetic field could be playing a role in
confining the expanding structures as they attempt to break out of the
disk.  The existence of ionized gas at heights of $\sim$~2.6~kpc above
the plane of the galaxy is particular puzzling.  If this gas is
photoionized, either directly or indirectly, by photons produced by massive
stars in the disk, then extremely long pathlengths through the halo are
required.

Our observations have allowed us to demonstrate quantitatively for the
first time that there is a continuous trend in [OII]/H$\alpha$+[NII]
and [SII]/H$\alpha$+[NII] as one proceeds to fainter and fainter
surface brightnesses.  This trend extends all the way from bright HII
region cores to filaments and loops and finally to faint diffuse
patches of emission, strongly suggestive that all are ionized by the
common process of photoionization.  The smooth sequence in ionization
state between HII regions and diffuse ionized gas revealed by our plots
lends strong support for photoionization of the diffuse gas by a dilute
radiation field, such as that produced by massive stars that are
located at considerable distances  (Mathis 1986, Domg\"orgen \& Mathis
1994).  In such a scenario, one could envisage the Lyman continuum
photons escaping from the HII regions where they were produced and
travelling significant distances through the ISM.   The propagation of
this radiation through the disk could be significantly altered by even
moderate columns of interstellar material, possibly resulting in a
hardened halo radiation field (e.g. Sokolowski 1994).
 While the smooth sequence in line ratios alone cannot be used to rule
 out shock ionization of the diffuse gas, the relatively small range in
observed values of these features makes a dominant role for shocks
rather unattractive, given the dependence on parameters such as shock
velocity.

The strong [OII] emission observed above the plane of the galaxy has
important implications for the electron temperature, $T_e$, of the
gas.  The [OII] line requires $\sim$ 3.3~eV to be collisionally excited
and the intensity of the line is governed by the factor
$T_{e}^{-0.5}~e^{-3.3~eV/kT_e}$ (Spitzer 1978).   The temperature
dependence  of the H$\alpha$ recombination line is approximately T$_{e}^{-1}$
(Osterbrock 1989).  Thus, as T$_e$ decreases from 10,000 K to 5,000K,
the strength of the [OII]/H$\alpha$ line ratio is expected to decrease
by a factor of $\sim$ 65.  Thus, the presence of widespread [OII]
emission above the plane of the galaxy requires warm gas in the
disk--halo interface.   The strength of the [OII] emission also places
severe constraints on the power source for the diffuse ionized gas.
Studies of the Milky Way and other galaxies have shown  that mechanical
energy from supernovae and stellar winds can barely satisfy the {\it
minimum} power requirements for diffuse ionized gas.  Strong,
ubiquitous [OII] emission implies that the minimum power model is
inappropriate for these calculations, suggesting only a minor role for
shocks in ionizing the diffuse gas.  The observed [OII]/H$\alpha$+[NII]
ratios in the extra--planar ionized features are larger than predicted
by the models of Domg\"orgen and Mathis (1994) for photoionization of
gas by a dilute radiation field.  This may suggest that a significant
hardening of the radiation does indeed take place as photons propagate
through the interstellar medium (see Sokolowski 1994).  Future
spectroscopic observations can extend the sensitivity to ionization
states through measurements of the [OIII] line and will also be
required to confirm the strength of the [OII]/H$\alpha$ emission
measurements in this and other galaxies.

The strong [OII] emission from the ionized gas in NGC~55 is of
particular interest in the interpretation of the redshifted emission
from more distant galaxies.  Much of the peculiar structure seen in
these galaxies (e.g. Schade et al 1996; Cowie et al 1995) may simply be
due to individual star--forming complexes and bright diffuse ionized
extensions, such as we have seen here.  Indeed, if NGC~55 were placed
at  z$\sim$ 0.6, it would very likely appear as a $`$chain' galaxy in
an HST F606W image (Cowie et al 1995). Further [OII] observations of
low redshift galaxies, spanning a range of morphological types and star
formation rates, will shed more light on the nature and evolutionary
state of moderate redshift galaxies.

We thank Tim Heckman, Piero Rosati and Deidre Hunter for stimulating
discussions and Don Cox for organizing an excellent Diffuse
Interstellar Medium session at the Madison, AAS meeting where we
presented preliminary results and enjoyed many discussions with
participants.  We thank the staff of Cerro Tololo Inter--American
Observatory for their excellent support.  AMNF acknowledges support
from the Zonta International Foundation in the form of an Amelia
Earhart Fellowship.  This research has been supported in part by NASA
grant NAGW--2892.
 
\pagebreak

 \pagebreak
\centerline{\bf FIGURES}

\noindent{Figure 1.~~ A mosaiced continuum--subtracted H$\alpha$ image
of NGC~55 displayed with a logarithmic stretch. 
The image shows the galaxy out to $\sim$ R$_{25}$.  North is up and east
is the left.}

\noindent{Figure 2(a).~~  A continuum--subtracted H$\alpha$ image of the
central 5~kpc of NGC~55, shown on a linear scale with individual features
marked.   (b) has been
smoothed with a gaussian with FWHM$=$ 2.5$''$ to enhance the low
surface brightness features.}

\noindent{Figure 3(a). ~~ The variation of [OII]/H$\alpha$+[NII] vs
H$\alpha$+[NII] surface brightness for various HII regions and diffuse
ionized gas features in NGC~55.  HII regions are marked by ($\ast$) and
diffuse ionized gas features by ($\Diamond$).  (b) ~~ The variation of
[SII]/H$\alpha$+[NII] vs H$\alpha$+[NII] surface brightness for various
HII regions and diffuse ionized gas features in NGC~55. }

\noindent{Figure 4(a) ~~The
variation of [OII]/H$\alpha$+[NII] vs z (height above the plane in
parsecs) in NGC~55. HII regions are marked by ($\ast$) and
diffuse ionized gas features by ($\Diamond$).  North and South are labelled.  (b) The variation of
[SII]/H$\alpha$+[NII] vs z (height above the plane in parsecs).}

\appendix
\clearpage

\end{document}